\documentstyle[aps,prb,preprint]{revtex}

\begin{document}
\author{M. I. Katsnelson$^{1}$ and A. V. Trefilov$^{2}$}
\address{$^{1}$Institute of Metal Physics, Ekaterinburg 620219, Russia\\
$^{2}$Russian Science Center Kurchatov Institute, Moscow 123182, Russia}
\title{Fermi liquid theory of electronic topological transitions and screening
anomalies in metals}
\date{\today}
\maketitle
\draft

\begin{abstract}
General expressions for the contributions of the Van Hove singularity (VHS)\
in the electron density of states to the thermodynamic potential $\Omega $
are obtained in the framework of a microscopic Fermi liquid theory. The
renormalization of the singularities in $\Omega $ connected with the
Lifshitz electronic topological transition (ETT) is found. Screening
anomalies due to virtual transitions between VHS and the Fermi level are
considered. It is shown that, in contrast with the one-particle picture of
ETT, the singularity in $\Omega $  turns out to be two-sided for interacting
electrons.
\end{abstract}


\narrowtext

Electronic topological transitions (ETT) \cite{lifshitz} can take place in
metals and alloys at the motion of the Van Hove singularities (VHS) of the
electron density of states across the Fermi level at the variation of
external parameters. Their investigation is now a well-developed branch of
solid state physics (see, e.g., the reviews \cite{blanter,KNT}). Recently
the interest in the problem has been revived by the observations of
anomalies in the pressure dependence of lattice properties of Cd and Zn,
which were explained in terms of ETT \cite{zncd}. To describe qualitatively
the behavior of thermodynamic and transport properties of metals near ETT,
correlation effects are to be taken into account since they are by no means
small at typical electron densities. Up to now the influence of the
interelectron interactions on ETT has been considered only for simplified
models. An exact expression for the many-electron renormalization of the
most singular contribution to the thermodynamic potential $\Omega $ near ETT
has been obtained in the model of nearly free electrons \cite{VT}. In
comparison with the one-particle expression for singular contribution to the
thermodynamic potential $\Omega ,$ correlation effects result in appearance
of numerical factors which are expressed in terms of the effective mass of
the quasiparticles and three-leg vertex $\gamma $ for the wave vectors
connecting VHS points in the Brillouin zone. Dzyaloshinskii \cite{dz}
demonstrated that in the two-dimensional (2D) case, in contrast with 3D one, 
$\gamma $ diverges and, moreover, the ground state of many-electron system
can be of a non-Fermi-liquid type provided that VHS are close enough to the
Fermi level (see also the recent papers \cite{twodim}). In the 3D case the
Fermi-liquid description is valid near ETT. However, the Fermi-liquid
renormalization factors contain singularities themselves, and some anomalous
contributions to the physical properties appear due to virtual transitions
between the VHS point in the electron energy spectrum and the Fermi level
(screening anomalies \cite{KT1}). It was found in \cite{KT2} that these
anomalies make the singularity in $\Omega $ at ETT to be two-sided, i.e., of
order of $\left( \pm z\right) ^{5/2}\theta \left( \pm z\right) -const\left(
\mp z\right) ^{7/2}\theta \left( \mp z\right) $ where $z=E_F-E_c$; $E_F$ and 
$E_c$ are the Fermi energy and the energy of VHS, $\theta \left( x>0\right)
=1,$ $\theta \left( x<0\right) =0.$ However, these considerations were based
on the calculations of particular diagrams in perturbation theory. In this
work we present a general consideration of the singularities in the
thermodynamic properties of metals near ETT in the framework of the rigorous
microscopic Fermi liquid theory \cite{migdal}. We restrict ourselves only to
the consideration of the 3D case and do not discuss the much more
complicated 2D case where the applicability of the Fermi liquid theory is
doubtful.

Let $\lambda =n{\bf k}\sigma $ be the set of electron quantum numbers in a
metal: band index, quasimomentum and spin projection, respectively. For the
normal Fermi liquid we have the following general expression for the Green
function near the chemical potential level $\mu $ \cite{migdal} 
\begin{eqnarray}
G_\lambda \left( E\right)  &=&\frac 1{E-\varepsilon _\lambda ^0+\mu -\Sigma
_\lambda \left( E,\mu \right) } \\
&=&\frac{Z_\lambda }{E+\mu -\varepsilon _\lambda }+G_\lambda ^{reg}\left(
E\right) ,  \nonumber  \label{1}
\end{eqnarray}
where $\varepsilon _\lambda ^0$ is the bare electron energy, $\varepsilon
_\lambda $ is the renormalized one that satisfies the equation

\begin{equation}
\varepsilon _\lambda =\varepsilon _\lambda ^0+\Sigma _\lambda \left(
\varepsilon _\lambda -\mu ,\mu \right) ,  \label{2}
\end{equation}
$\Sigma _\lambda \left( E,\mu \right) $ is the self-energy, $%
\mathop{\rm Im}
\Sigma _\lambda \left( E=0,\mu \right) =0,$ $Z_\lambda $ is the residue of
the Green function in the pole and $G_\lambda ^{reg}\left( E\right) $ is the
regular (incoherent or nonquasiparticle) part of the Green function. To find
the singular contribution to the thermodynamic potential connected with the
closeness of VHS in the {\it renormalized} spectrum to the Fermi energy $%
E_F=\mu $ it is suitable to start from the Luttinger theorem \cite{luttinger}
\begin{equation}
N=-\frac{\partial \Omega }{\partial \mu }=\sum\limits_\lambda \theta \left(
\mu -\varepsilon _\lambda \right)   \label{3}
\end{equation}
where $N$ is the number of particles, 
\[
\sum\limits_\lambda =\sum\limits_{n\sigma }\frac{V_0}{\left( 2\pi \right) ^3}%
\int\limits_{BZ}{\bf dk}
\]
BZ is the Brillouin zone, $V_0$ is the unit cell volume. Differentiating (3)
with respect to $\mu $ with taking into account Eq. (1) one has 
\begin{eqnarray}
\frac{\partial N}{\partial \mu } &=&-\frac{\partial ^2\Omega }{\partial \mu
^2}=\sum\limits_\lambda \delta \left( \mu -\varepsilon _\lambda \right) 
\frac{\partial \left( \mu -\varepsilon _\lambda \right) }{\partial \mu } \\
&=&\sum\limits_\lambda \delta \left( \mu -\varepsilon _\lambda \right) \frac{%
\partial G_\lambda ^{-1}\left( E=0,\mu \right) }{\partial \mu }Z_\lambda  
\nonumber  \label{4}
\end{eqnarray}
Further, we use the set of well-known identities \cite{migdal}. First of
all, the quantity 
\begin{equation}
\frac{\partial G_\lambda ^{-1}\left( E=0,\mu \right) }{\partial \mu }=\gamma
_{\lambda \lambda }  \label{5}
\end{equation}
is the three-leg vertex describing the response to the uniform static field.
It is connected with the ``dynamic'' response $\gamma _{\lambda \lambda
}^\omega $ by the equation 
\begin{equation}
\gamma _{\lambda \lambda }=\gamma _{\lambda \lambda }^\omega
-\sum\limits_\nu \Gamma _{\lambda \nu }^\omega Z_\nu ^2\delta \left( \mu
-\varepsilon _\nu \right) \gamma _{\nu \nu }  \label{6}
\end{equation}
where $\Gamma _{\lambda \nu }^\omega $ is the ``dynamic'' limit of the
four-leg vertex which is connected with the Landau Fermi-liquid interaction
function 
\begin{equation}
f_{\lambda \nu }=\frac{\delta ^2E_{tot}}{\delta n_\lambda \delta n_\nu }
\label{7}
\end{equation}
($E_{tot\text{ }}$ is the total energy, $n_\lambda $ is the quasiparticle
distribution function) by the relation 
\begin{equation}
f_{\lambda \nu }=Z_\lambda Z_\nu \Gamma _{\lambda \nu }^\omega   \label{8}
\end{equation}
At the same time, the Ward identity gives 
\begin{equation}
\gamma _{\lambda \lambda }^\omega =\frac 1{Z_\lambda }  \label{9}
\end{equation}
On substituting Eqs. (5-9) into (4) we derive the following exact expression 
\begin{equation}
\frac{\partial ^2\Omega }{\partial \mu ^2}=-\sum\limits_\lambda \delta
\left( \mu -\varepsilon _\lambda \right) \widetilde{\gamma }_\lambda 
\label{10}
\end{equation}
where $\widetilde{\gamma }_\lambda =\gamma _{\lambda \lambda }Z_\lambda $
satisfies the equation 
\begin{equation}
\widetilde{\gamma }_\lambda =1-\sum\limits_\nu f_{\lambda \nu }\delta \left(
\mu -\varepsilon _\nu \right) \widetilde{\gamma }_\nu   \label{11}
\end{equation}
It follows from Eq. (10) that the singular contribution to $-$ $\partial
^2\Omega /\partial \mu ^2$ is proportional to the singularity in the density
of states at the Fermi level, 
\begin{equation}
\rho \left( \mu \right) =\sum\limits_\lambda \delta \left( \mu -\varepsilon
_\lambda \right)   \label{12}
\end{equation}
The coefficient of the proportionality may be found from the solution of the
integral equation (11) for a given quasiparticle spectrum and interaction
function.

This form of the result is probably most convenient to separate the
singularity of order of $\left( \pm z\right) ^{5/2}\theta \left( \pm
z\right) $. To investigate the screening anomalies it is better to use
another expression which may be obtained directly from Eqs. (1), (4) 
\begin{equation}
-\frac{\partial ^2\Omega }{\partial \mu ^2}=\sum\limits_\lambda Z_\lambda
\delta \left( \mu -\varepsilon _\lambda \right) \left[ 1-\frac{\partial
\Sigma _\lambda \left( E=0,\mu \right) }{\partial \mu }\right]   \label{13}
\end{equation}
One can see from the perturbation expansion \cite{KT1,KT2} that the
corresponding contributions to the $\lambda $-dependence of $\Sigma $ is
weaker than to its energy dependence and can be neglected when separating
the main singularity. Thus the multiplier $Z_\lambda \delta \left( \mu
-\varepsilon _\lambda \right) $ in Eq. (13) turns out to be nonsingular, and
we have to consider only the term with $\partial \Sigma /\partial \mu $.

The second-order expression for the self-energy has the form (see e.g. \cite
{lich}) 
\begin{eqnarray}
\Sigma _\lambda ^{(2)}\left( E=0,\mu \right)  &=&\sum\limits_{\left\{
\lambda _i\right\} }U_{\lambda \lambda _1\lambda _3\lambda _2}^dU_{\lambda
_1\lambda _2\lambda \lambda _3}  \label{14} \\
&&\ \times \frac{\left[ n_{\lambda _3}\left( 1-n_{\lambda _1}-n_{\lambda
_2}\right) +n_{\lambda _1}n_{\lambda _2}\right] }{\mu +\varepsilon _{\lambda
_3}-\varepsilon _{\lambda _1}-\varepsilon _{\lambda _2}}  \nonumber
\label{14}
\end{eqnarray}
where 
\begin{eqnarray}
U_{\lambda _1\lambda _2\lambda _4\lambda _3} &=&\left\langle \lambda
_1\lambda _2\left| U\right| \lambda _4\lambda _3\right\rangle   \label{15} \\
U_{\lambda \lambda _1\lambda _3\lambda _2}^d &=&2U_{\lambda \lambda
_3\lambda _1\lambda _2}-U_{\lambda \lambda _3\lambda _2\lambda _1}, 
\nonumber
\end{eqnarray}
$U$ is the interelectron interaction. Here $\lambda _i$ are the {\it orbital}
quantum numbers, and we restrict ourselves to the nonmagnetic case. In
particular, in one-band Hubbard model one has \widetext
\begin{equation}
\Sigma _{{\bf k}}^{(2)}\left( E=0,\mu \right) =U^2\sum\limits_{{\bf k}_1{\bf %
k}_2}\frac{\left[ n_{_{{\bf k}_1{\bf -k}_2}}\left( 1-n_{{\bf k}_1}-n_{{\bf k}%
_2}\right) +n_{{\bf k}_1}n_{{\bf k}_2}\right] }{\mu +\varepsilon _{{\bf k}_1%
{\bf -k}_2}-\varepsilon _{{\bf k}_1}-\varepsilon _{{\bf k}_2}}  \label{16}
\end{equation}
\narrowtext
Averaging in ${\bf k}$ over the Brillouin zone we find after simple
transformations 
\begin{equation}
\sum\limits_{{\bf k}}\frac{\partial ^2\Sigma _{{\bf k}}^{(2)}\left( E=0,\mu
\right) }{\partial \mu ^2}=\frac{3U^2\rho ^2}8R\left( \mu \right) 
\label{17}
\end{equation}
where 
\[
R(z)=P\int \frac{d\varepsilon \rho \left( \varepsilon \right) }{%
z-\varepsilon }
\]
is the real part of the on-site lattice Green function. Thus one obtains for
the contribution $\Omega _{sa}$ of screening anomalies to $\Omega $%
-potential near ETT $\partial ^3\Omega _{sa}/\partial \mu ^3\propto
U^2R\left( \mu \right) .$ On taking into account that the singularity of
order of $\left( \pm z\right) ^{1/2}\theta \left( \pm z\right) $ in $\rho
\left( \mu \right) $ corresponds to the singularity of order of $\left( \mp
z\right) ^{1/2}\theta \left( \mp z\right) $ in $R\left( \mu \right) ,$ we
conclude from Eqs. (13), (17) that the singularity in $\Omega \left( \mu
\right) $ is two-sided, $\delta \Omega \left( \mu \right) \propto \left( \pm
z\right) ^{5/2}\theta \left( \pm z\right) -const\left( \mp z\right)
^{7/2}\theta \left( \mp z\right) ,$ as it was mentioned above.

General equations (10), (11) seem to be formally applicable also for
low-dimensional systems with a non-Fermi-liquid ground state. However, in
these cases the Landau function $f_{\lambda \nu }$ is divergent for the
forward-scattering processes \cite{anderson}. We will not consider here this
complicated case.

To conclude, it is worthwhile to note one more consequence of these
equations. There are two types of quasiparticles in interacting Fermi
systems: dynamical quasiparticles with the spectrum determined by the poles
of Green functions and statistical quasiparticles, i.e. quasiparticles in
the sense of Landau theory \cite{noz}. Generally speaking, their spectra do
not coincide (the difference  occurs in the third order of perturbation
expansion for paramagnetic state \cite{pethick} and in the second order for
ferromagnetic state \cite{ufn}). However, we prove that the points of ETT
found from both spectra are the same since $\rho \left( \mu \right) $is
determined by the dynamical quasiparticles whereas $\Omega \left( \mu
\right) $by statistical ones. Together with the Luttinger theorem this means
that both the volume and topology (but not necessarily the exact shape) of
the Fermi surface are the same for these two types of quasiparticles.

\end{document}